\documentclass{WileyMSP-template}

\usepackage{ragged2e}
\usepackage{siunitx}
\usepackage{booktabs}

\begin{document}

\pagestyle{fancy}
\rhead{\includegraphics[width=2.5cm]{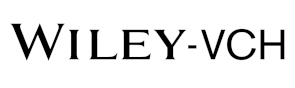}}

\title{Broadband parametric amplification in AlGaAs-on-insulator \\
nanowaveguides}

\maketitle


\author{Yanjing Zhao}
\author{Chanju Kim}
\author{Yi Zheng}
\author{Chaochao Ye}
\author{Yueguang Zhou}
\author{Kresten Yvind}
\author{Minhao Pu*}


\begin{affiliations}
Y. Zhao, C. Kim, Y. Zheng, C. Ye, Y. Zhou, K. Yvind, and M. Pu\\
DTU Electro, Department of Electrical and Photonics Engineering, Technical University of Denmark, 2800 Kgs. Lyngby, Denmark\\
Y. Zhao\\
Present Address: National Information Optoelectronics Innovation Center, Wuhan 430074, China\\
Email address: mipu@dtu.dk

\end{affiliations}


\keywords{nonlinear optics, four-wave-mixing, parametric amplifier, integrated optics}

\justifying

\begin{abstract}

Optical amplification is critical for optical signal transmission. While the emergence of erbium-doped fiber amplifiers has revolutionized optical communications in fiber-based systems, on-chip amplification is still needed for integrated optics. Since nanoscale waveguides enhance nonlinearity by several orders of magnitude, they are promising candidates for optical parametric amplification. Using a pulsed pump at 1550 nm, broadband optical parametric amplification based on four-wave mixing is investigated in AlGaAs-on-insulator nanowaveguides. The intense nonlinearity enables an on--off gain as large as 58.4 dB. Meanwhile, the low propagation loss leads to a 56.2-dB net on-chip gain. Combined with further dispersion engineering, the net on-chip gain bandwidth extends beyond 415 nm, which is 2.3 times larger than previous reports pumped in the telecom band in integrated optics. The demonstrated results represent the largest parametric gain and bandwidth reported for on-chip parametric amplifiers.

\end{abstract}


\section{Introduction}
Optical amplification is generally pursued in integrated photonics for communications, ultrafast spectroscopy, and sensing \cite{hansryd2002fiber}. Historically, semiconductor optical amplifiers (SOAs) based on stimulated emission in III--V gain media emerged as the first integrated-compatible amplifiers, offering compact, electrically driven gain \cite{olsson2002lightwave,sobhanan2022semiconductor}. However, they suffer from a high amplified spontaneous emission (ASE) noise figure and carrier-induced saturation \cite{sobhanan2022semiconductor}. Erbium-doped waveguide amplifiers (EDWAs) provide low-noise gain and stable operation, yet their gain bandwidth is intrinsically restricted by discrete energy levels \cite{bradley2011erbium,liu2022photonic}. Nonlinear scattering-based mechanisms, including stimulated Raman \cite{rong2007low} and Brillouin \cite{kittlaus2016large} amplification, have also been explored on-chip; however, their gain bandwidths are intrinsically narrow, constrained by the phonon resonance linewidth of the host material. In contrast, optical parametric amplification (OPA) based on four-wave mixing (FWM) offers a compelling alternative: its gain bandwidth is governed by the phase-matching condition rather than a fixed material transition, enabling broadband amplification that can be flexibly tailored through dispersion engineering, while its noise performance can approach the fundamental quantum limit in the phase-sensitive regime.\\
Early research was generally carried out in fibers \cite{ho2001200,torounidis2006fiber,xing2017mid,wu2019net,wu2020four}, demonstrating a 70-dB on--off gain \cite{torounidis2006fiber} and a 200-nm net on--off gain bandwidth \cite{ho2001200}. Subsequently, nanoscale waveguides emerged with optical nonlinearities increased by several orders of magnitude, enabling integrated chip-scale amplifiers. Optical parametric amplification has been demonstrated in various platforms, such as silicon (Si) \cite{zlatanovic2010mid,foster2006broad,liu2012bridging,kuyken201150,liu2010mid}, amorphous silicon (a-Si) \cite{kuyken2011chip}, silicon nitride ($\rm Si_{3}N_4$) \cite{lupken2021optical,ye2021overcoming}, ultra-silicon-rich nitride ($\rm Si_{7}N_3$) \cite{choi2018broadband,sahin2020optical,ooi2017pushing}, and chalcogenides (ChG) \cite{lamont2008net}. Although a large net on-chip gain of up to 42.3 dB has been achieved in silicon waveguides by pumping in the mid-infrared \cite{kuyken201150}, it has been limited to 1.8 dB for signals in the telecom band due to strong nonlinear loss induced by two-photon absorption (TPA) \cite{foster2006broad}. By modifying the material properties to enhance nonlinearity, a 21.2-dB net on-chip gain has been obtained in amorphous silicon pumped at 1535 nm \cite{kuyken2011chip}. In addition, free from the influence of TPA, the on--off gain has been increased to 42.5 dB in the telecom band in ultra-silicon-rich nitride \cite{ooi2017pushing}. A 32.5-dB net on-chip gain has also been demonstrated in chalcogenide glasses \cite{lamont2008net}. Beyond the demonstrations above using pulsed pumping, a 6.4-dB on--off gain has been obtained in a 1.4-m-long silicon nitride waveguide pumped by a 1563-nm continuous-wave (CW) laser \cite{ye2021overcoming}.
Further progress includes 25-dB net gain in the gallium phosphide-on-silicon dioxide platform \cite{kuznetsov2025ultra} and broadband amplification in the silicon nitride platform \cite{zhao2025ultra}. \\
Since FWM efficiency is related to phase matching, nonlinear intensity, and loss, aluminum gallium arsenide (AlGaAs), with strong nonlinearity \cite{aitchison1997nonlinear}, emerges as a promising candidate to further improve optical parametric gain \cite{kultavewuti2015low}. Compared with the materials above, AlGaAs possesses a high nonlinear index of $10^{-17}$ $\rm m^2/W$, two orders of magnitude larger than that of silicon nitride \cite{ji2017ultra}. Importantly, its material bandgap can be engineered to suppress TPA in the telecom region \cite{wathen2014efficient}. Combined with a linear refractive index of 3.3, the AlGaAs-on-insulator (AlGaAsOI) platform \cite{pu2016efficient} enables strong light confinement and a large effective nonlinear coefficient, reaching up to 720 $\rm W^{-1}m^{-1}$ \cite{stassen2019ultra}. This large nonlinearity, together with the high index contrast, facilitates both efficient parametric processes and flexible dispersion engineering for broadband operation \cite{pu2018integrated, kong2022super}. A wide range of nonlinear processes including soliton comb generation \cite{moille2020dissipative, chen2024integrated, zhao2025thermal}, supercontinuum generation \cite{kuyken2020octave,may2021supercontinuum}, second-harmonic generation \cite{may2019second}, and parametric down-conversion \cite{placke2024telecom} have been demonstrated in AlGaAsOI, highlighting its versatility as a nonlinear platform. These features make AlGaAsOI an attractive candidate for broadband optical parametric amplification, which has not yet been explored.

\begin{figure*}[!htb]
\centering
\includegraphics[width=\linewidth]{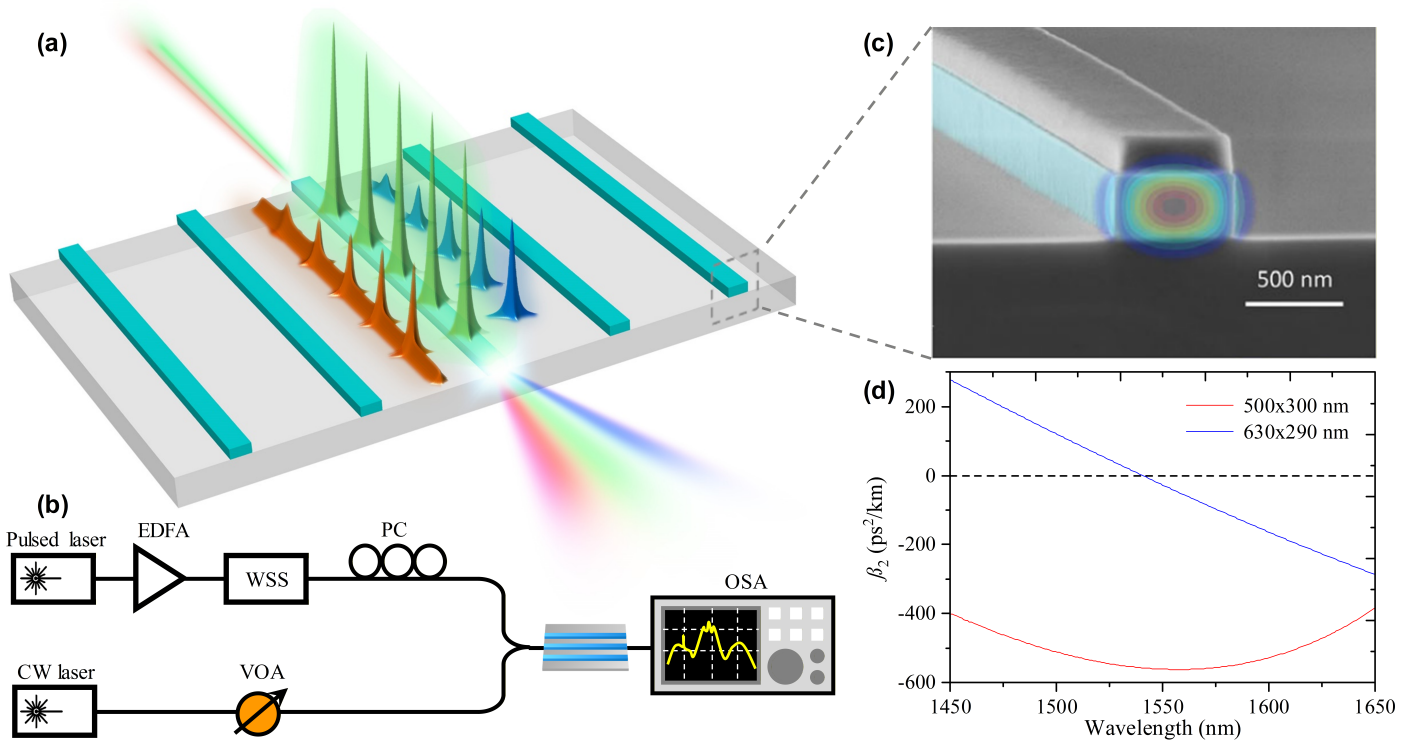}
\captionsetup{labelformat=default,labelsep=period} 
\caption{Nonlinear photonic chip for optical parametric amplification. (a) Schematic illustration of FWM-based parametric amplification. (b) Experimental setup of FWM-based parametric amplification. EDFA: erbium-doped fiber amplifier; WSS: wavelength selective switch; PC: polarization controller; VOA: variable optical attenuator; OSA: optical spectrum analyzer. (c) SEM picture of an AlGaAsOI nano-waveguide. The thin ridge cladding on top corresponds to the residual hydrogen silsesquioxane (HSQ) resist remaining after the etching process, originally used as a hard mask. (d) Dispersion of waveguides with different cross-sections.}
\label{figure1}
\end{figure*}

In this work, we utilize an AlGaAsOI waveguide to explore optical parametric amplification based on FWM, using a picosecond pulsed pump in the telecom band. Benefiting from the high nonlinearity and low loss, we demonstrate on--off signal parametric gain and idler translation gain of up to 56.9 dB and 58.4 dB, respectively. Moreover, owing to the low propagation loss, the net on-chip gains for the signal and idler are as large as 54.7 dB and 56.2 dB. With further dispersion engineering, the net on-chip gain bandwidth exceeds 415 nm. These results represent the largest gain and amplification bandwidth pumped in the telecom band in integrated optics to date.

\section{Methods}
The schematic illustration of FWM-based optical parametric amplification is presented in Figure~\ref{figure1}a. The pump and signal are injected into the nonlinear waveguide. In the FWM process, two pump photons at frequency $\omega_p$ are converted into signal and idler photons at frequencies $\omega_s$ and $\omega_i$, respectively, and energy conservation requires $2\omega_{p} = \omega_{s} + \omega_{i}$. As the pump photons are annihilated, the signal is amplified and the idler is generated. The parametric gain is determined by the FWM efficiency, which is related to phase mismatch induced by dispersion and nonlinear effects \cite{boyd2020nonlinear}:
\begin{equation}
\Delta{k} = 2{\gamma}P_{p} - \Delta{k}_L
\label{eq:1}
\end{equation}
Here, $\gamma$ is the effective nonlinear coefficient of the waveguide, $P_p$ is the pump peak power, and $\Delta{k_L} = 2{k_p} - {k_s} - {k_i}$ is the linear phase mismatch, where $k_p$, $k_s$, and $k_i$ correspond to the pump, signal, and idler wavevectors, respectively. Considering only group-velocity dispersion (GVD) $\beta_2$ and ignoring higher-order dispersion, the linear phase mismatch can be simplified as $\Delta{k_L} \approx -{\beta_2}\Delta\omega^2$, where $\Delta\omega = \omega_{p} - \omega_{s}$ is the frequency difference between the pump and signal. Guided by these parameters, the FWM gain coefficient $g$ is given by:
\begin{equation}
g = \sqrt{{\gamma}P_{p}\Delta{k}_L - \left(\frac{\Delta{k}_L}{2}\right)^2}
\label{eq:2}
\end{equation}
The signal gain is then defined as
\begin{equation}
G_s = \frac{P^{out}_s}{P^{in}_s} = 1 + \left[\frac{{\gamma}P_{p}}{g} \cdot \sinh\left(gL\right)\right]^2 
= 1 + \left({\gamma}P_{p}L\right)^2 + \left(1 + \frac{{g^2}{L^2}}{6} + \frac{{g^4}{L^4}}{120} + \ldots\right)
\label{eq:3}
\end{equation}
where $P^{in}_s$ and $P^{out}_s$ are the input and output powers of the signal, $L$ is the waveguide length, and the last equality is derived from the Taylor expansion of $\sinh(x)$. At the phase-matching point ($\Delta{k} = 0$), the peak gain is obtained and can be approximated as $\exp(2\gamma{P_p}L)/4$\cite{hansryd2002fiber}. Therefore, large nonlinearity, high pump power, and a long interaction length are desired for high parametric gain. On the other hand, to achieve phase matching, anomalous dispersion ($\beta_2<0$) is required to compensate the positive nonlinear phase mismatch $2\gamma{P_p}$. With a larger absolute value of GVD, the phase mismatch increases sharply as the frequency moves away from the phase-matching point, leading to a rapid decrease in parametric gain. Thus, a small absolute GVD is preferred for broadband amplification, as it can limit the phase mismatch over a wide spectral range. In addition, a longer interaction length $L$ amplifies the impact of phase mismatch under imperfect phase-matching conditions, leading to a more rapid spectral roll-off and thus reduced gain flatness across the amplification bandwidth. (see Supporting Information). Consequently, intense nonlinearity and near-zero anomalous dispersion are desired for a large gain bandwidth.

The experimental setup is shown in Figure~\ref{figure1}b. The pulsed pump and CW signal are combined with a 50:50 coupler and then injected into the AlGaAsOI waveguide via a lensed fiber. The pump originates from a femtosecond source centered at 1550 nm with a 90-MHz repetition rate. It is reshaped into a Gaussian spectrum with 4.8-nm bandwidth, and dispersion is compensated using a wavelength selective switch (WSS), resulting in a pulse duration of 1 ps. The signal is provided by a CW tunable laser covering 1480-1630 nm range. Polarization controllers (PCs) are used to co-polarize the pump and signal for efficient excitation of the fundamental TE mode. The output is then directed to an optical spectrum analyzer (OSA) for spectral measurement.

To investigate optical parametric amplification in the AlGaAsOI platform, we use an 11-mm-long waveguide fabricated on an insulator with a width of 500 nm, a height of 300 nm, and a silica cladding thickness of \SI{3}{\micro\meter}. The AlGaAs layer has a 21\% aluminum composition, corresponding to a bandgap of $\sim1.72$ eV. The fabrication process is detailed in Ref.~\cite{ottaviano2016low, kim2022design}, and a scanning electron microscopy (SEM) image of the etched waveguide is shown in Figure~\ref{figure1}c. Due to the large index contrast between AlGaAs and silica, light is well confined in the waveguide and the dispersion can be efficiently tuned. Based on finite-element simulations, the red curve in Figure~\ref{figure1}d shows the GVD profile of the fundamental TE mode in a waveguide with a $500\times300$ nm cross section (Device 1). The anomalous dispersion spans 1450--1650 nm, including the pump wavelength (1550 nm). The GVD $\beta_2$ is $-561.5~\rm ps^2/km$ at the pump wavelength. The simulated GVD profile for a waveguide with a $630\times290$ nm cross section (Device 2) is also shown in Figure~\ref{figure1}d (blue curve). With a larger width and smaller height, the absolute GVD value is reduced to $-27.2~\rm ps^2/km$ at 1550 nm. The influence of dispersion on the net on-chip gain bandwidth will be discussed later; here we begin with the characterization of the $500\times300$ nm waveguide. Over this wavelength range, a low propagation loss of $2~\rm dB/cm$ is measured using the cut-back method.

\section{Results} 
\begin{figure}[!htb]
\centering
\includegraphics[width=12cm]{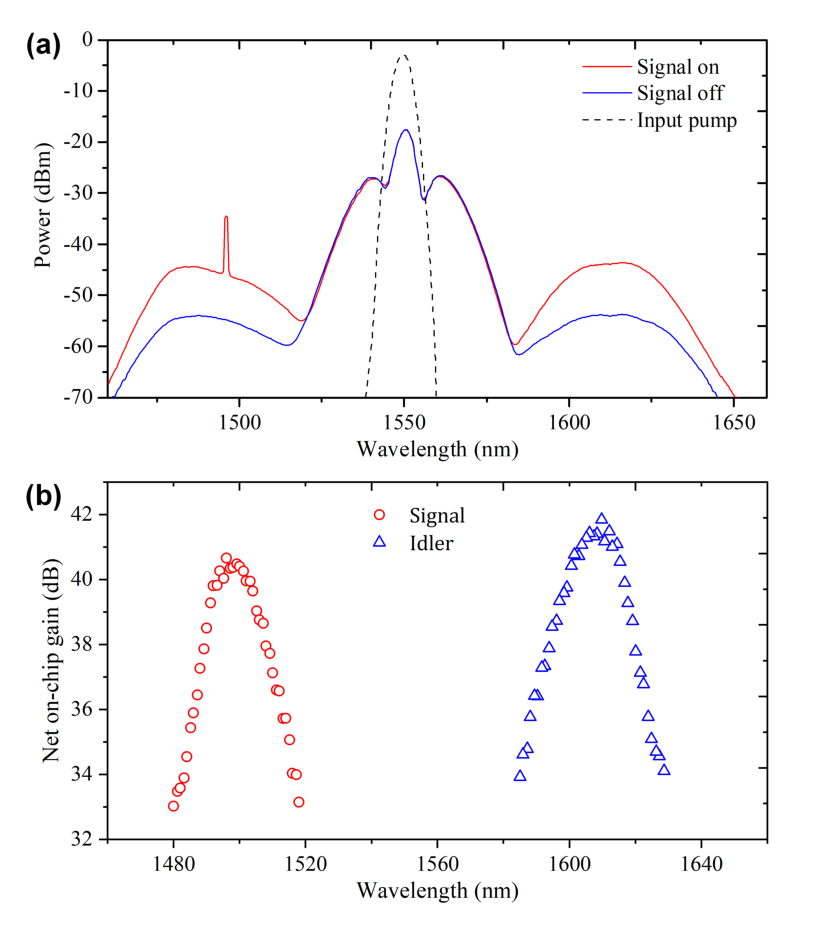}
\captionsetup{labelformat=default,labelsep=period} 
\caption{Spectra and net on-chip gain for optical parametric amplification. (a) Spectra with the signal on and off, and the input pump spectrum. (b) Net on-chip signal parametric gain (red circles) and idler translation gain (blue triangles).}
\label{figure2}
\end{figure}

First, to investigate broadband optical parametric amplification, the signal wavelength is tuned from 1480 to 1520 nm (1-nm step size). During the procedure, the peak pump power and the signal power at the waveguide input are fixed at 28.8 W and \SI{5.5}{\micro\watt}, respectively. The peak pump power is chosen such that the system operates in the small-signal (undepleted-pump) regime, where pump depletion remains negligible. This preserves phase matching along the waveguide, maintains the gain spectrum, and ensures a linear and predictable gain response for reliable extraction \cite{hansryd2002fiber}, while also supporting high signal fidelity in communication and signal processing applications \cite{marhic2015fiber}. Through the FWM process, energy is transferred from the high-power pump to the signal and idler, resulting in signal amplification and idler generation. The idler wavelength is dictated by energy conservation. With the signal tuned from 1480 to 1520 nm, the idler wavelength varies from 1627 to 1581 nm. The spectra with the signal on and off are recorded in Figure~\ref{figure2}a. Compared with the input spectrum, the output spectra are broadened by strong Kerr nonlinearities in the AlGaAsOI waveguide, where high-peak-power pump pulses induce pronounced self-phase modulation. In addition, a continuous noise background arising from noise-seeded spontaneous modulational instability is observed, reflecting the amplification of vacuum fluctuations under anomalous dispersion and serving as a qualitative indicator of the strong on-chip parametric gain, consistent with Fig.~\ref{figure2}b. Due to the pulsed nature of the pump, the generated signal and idler are also pulsed, with FWHMs similar to that of the pump. According to the analysis method in Ref.~\cite{liu2010mid}, the spectra in Figure~\ref{figure2}a are used to extract the net on-chip signal parametric gain (red circles) and idler translation gain (blue triangles), as shown in Figure~\ref{figure2}b. Here, the net on-chip gain is defined as the ratio between the peak power of the pulsed signal/idler at the waveguide output and the coupled CW signal power at the waveguide input. Over this tuning range, the AlGaAsOI waveguide exhibits excellent optical parametric amplification performance, with a net on-chip signal parametric gain of 40.7 dB and an idler translation gain of 41.8 dB. Moreover, the net on-chip gain is much larger than the chip-to-fiber coupling loss (approximately 2.3 dB), corresponding to a net off-chip gain of more than 31 dB for both the signal and idler from 1480 to 1520 nm.
\begin{figure}[!htb]
\centering
\includegraphics[width=10.5cm]{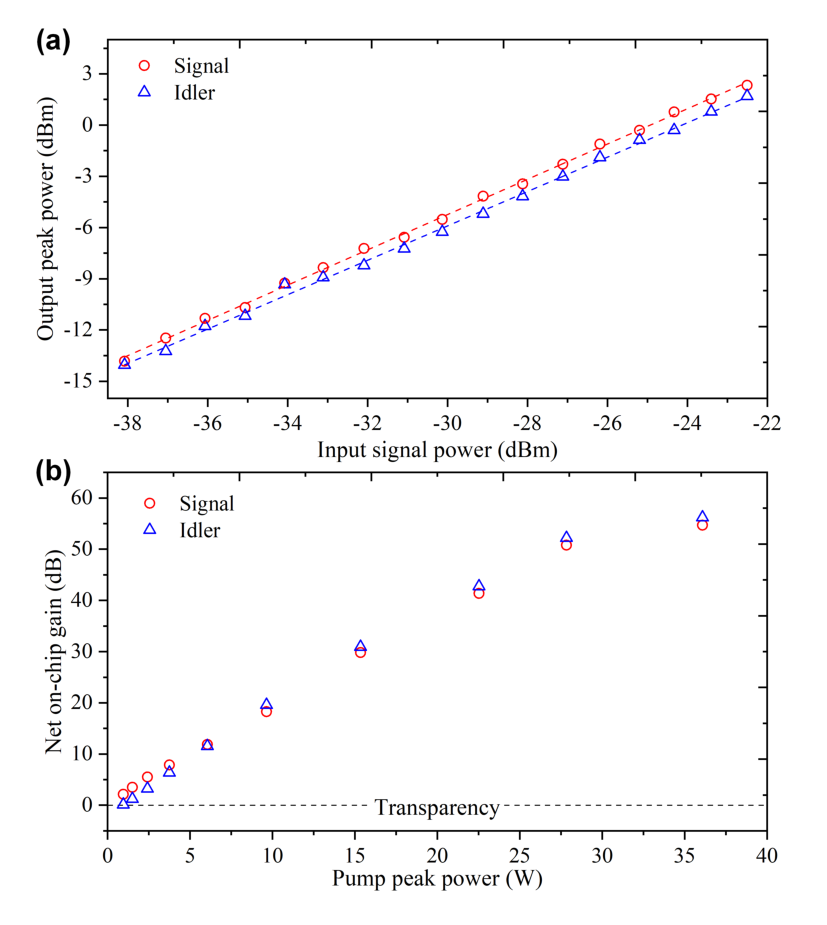}
\captionsetup{labelformat=default,labelsep=period} 
\caption{Performance characteristics of optical parametric amplification. (a) Output peak power of signal and idler pulses versus input signal power. (b) Net on-chip parametric and translation gain versus the peak power of the input pump pulse.}
\label{figure3}
\end{figure}

Next, the performance characteristics of optical parametric amplification are investigated. By fixing the input signal wavelength at the gain peak (1496 nm) and tuning the signal power at a pump peak power of 22.9 W, the output peak powers of the signal (red circles) and idler (blue triangles) are plotted in Figure~\ref{figure3}a. The optical parametric amplification is linear over a 16-dB dynamic range, as demonstrated by the dashed linear fits. Under these conditions, the corresponding net on-chip gains for the signal and idler are 34.8 dB and 34.5 dB, respectively.

On the other hand, by tuning the pump power, we sweep the input signal wavelength to extract the gain peak at each pump power, as shown in Figure~\ref{figure3}b. Notably, the input signal power is chosen to suppress the generation of cascaded FWM sidebands. The net on-chip gain increases significantly with pump power, reaching maxima of 54.7 dB and 56.2 dB for the signal and idler, respectively, which is 24.5 times larger than that previously reported in an 11-mm ultra-silicon-rich nitride waveguide\cite{ooi2017pushing}. Compared with the previous report, the smaller fiber-to-chip loss in our device leads to a larger on-chip pump power. Combined with a larger effective nonlinear coefficient, such intense parametric gain can be obtained. As indicated by the intersection with the dashed line in Figure~\ref{figure3}b, the on-chip gain threshold is 0.9 W. While the on-chip gain is slightly saturated at the maximum input pump power, it remains approximately linear for pump peak powers below 28 W, where both signal and idler gains exceed 50 dB. The saturation is induced by nonlinear loss, which might be caused by surface states\cite{grillanda2015light,Kamel2023SurfaceWaveguides}. Moreover, benefiting from the small coupling loss (approximately 2.3 dB), the off-chip signal parametric gain and idler translation gain are as large as 52.4 dB and 53.9 dB, respectively, which is 141 times larger than the previous best result\cite{ooi2017pushing}.
\begin{figure}[!htb]
\centering
\includegraphics[width=10.5cm]{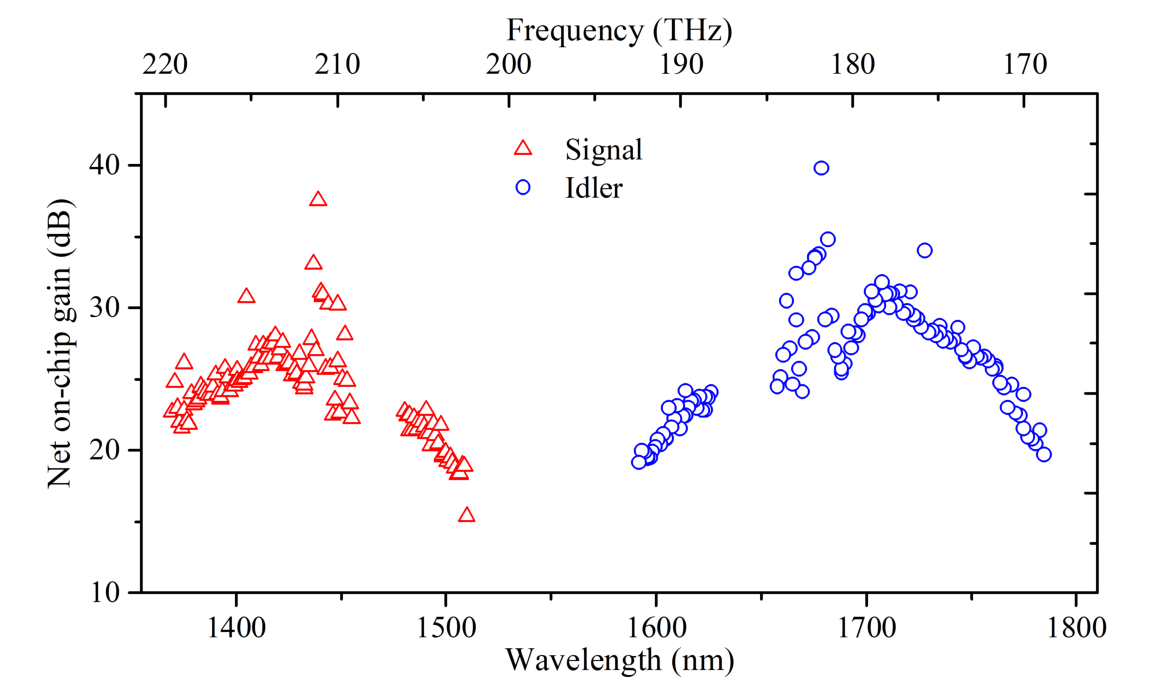}
\captionsetup{labelformat=default,labelsep=period} 
\caption{Optical parametric amplification for the waveguide with a $630\times290$ nm cross section.}
\label{figure4}
\end{figure}

Finally, optical parametric amplification is studied under a different dispersion condition. Here, we use a waveguide with a $630\times290$ nm cross section (Device 2), for which the GVD is as small as $-27.2~\rm ps^2/km$ at the pump wavelength (1550 nm). To investigate broadband parametric amplification, the tuning range of the signal wavelength covers 1370--1455 nm as well as 1480--1520 nm. The pump peak power is fixed at 16.6 W and the signal power is kept below 0.2 mW. 
Theoretically, as the absolute value of anomalous dispersion decreases, the gain peak moves farther away from the pump and the gain bandwidth increases, as illustrated in Figure~\ref{figure4}. Here, the red triangles and blue circles represent the net on-chip gains for the signal and idler, respectively. Notably, the net on-chip gain is larger than 15 dB over the entire tuning range, which spans about 415 nm from 1370 nm to 1785 nm. Limited by the tuning range of the CW lasers available in our lab, the full net on-chip gain bandwidth cannot be measured experimentally and is expected to exceed 415 nm. As presented in Table~\ref{Table1}, the net on-chip gain bandwidth is 2.3 times larger than the previous result in the chalcogenide platform \cite{lamont2008net}. In addition, the gain profile deviates slightly near 1680 nm, which may be induced by interference with the Raman peak \cite{kuyken201150}. It is worth noting that the lower peak gain observed here compared to Device 1 arises from the reduced pump peak power. To enable broadband measurement, a higher CW signal power was used to maintain sufficient signal-to-noise ratio across the extended tuning range (1370--1455 nm). Further increasing the pump power would induce cascaded four-wave mixing sidebands and spectral distortion, compromising accurate gain extraction; thus, the pump power was limited, resulting in reduced gain since it scales with pump peak power. In contrast, the amplification bandwidth is governed by dispersion engineering. A direct gain bandwidth comparison of the two devices at the same pump peak power (see Supporting Information) further confirms that dispersion is the governing parameter for the gain bandwidth.

\begin{table} [!ht]
\caption{Comparison of integrated optical parametric amplification parameters pumping at the telecom band.}
\label{Table1}
\begin{tabular}{rrrrrrl}
\toprule
Platform & $\rm \sf{n_{2} (m^2/W)}$ & Length & $\rm \sf{\beta_{2} (ps^{2}/km)}$ & Net on-chip gain (dB) & Bandwidth (THz) & Ref \\ \midrule
Si				& $4\sim9\times10^{-18}$     & 17 mm	   & -637  & 2.8  &	5.9	  & \cite{foster2006broad} \\
a-Si				& $1.1\sim1.5\times10^{-17}$ & 11 mm 	   & -2600 & 21.2 &	$>$9.7  &	\cite{kuyken2011chip} \\
$\rm {\sf Si_{7}N_{3}}$	& $2.8\times10^{-17}$	     & 7 mm		   & 280   & 39.4 &	$>$2.3  & \cite{ooi2017pushing} \\
$\rm \sf Si_{3}N_{4}$	& $2.4\times10^{-19}$     	 & 560 mm	   & -2.2   & 1  &	32.7	  & \cite{zhao2025ultra} \\
ChG  	& $3\times10^{-18}$	         & 65 mm	   & -37.4 & 32.5 &	$>$22.9 & \cite{lamont2008net} \\
GaP  	& $1.1\times10^{-17}$	         & 55.5 mm	   & -124 & 25 &	17 & \cite{kuznetsov2025ultra} \\
AlGaAs			& $10^{-17}$	             & 11 mm	   & -400  & 56.2 &	$>$18.5 & this work \\
AlGaAs			& $10^{-17}$            	 & 9 mm		   & -27.2 & 31.8 & $>$54.9  &	this work \\
\bottomrule
\end{tabular}
\end{table}

\section{Discussion}
\begin{figure}[!htb]
\centering
\includegraphics[width=14cm]{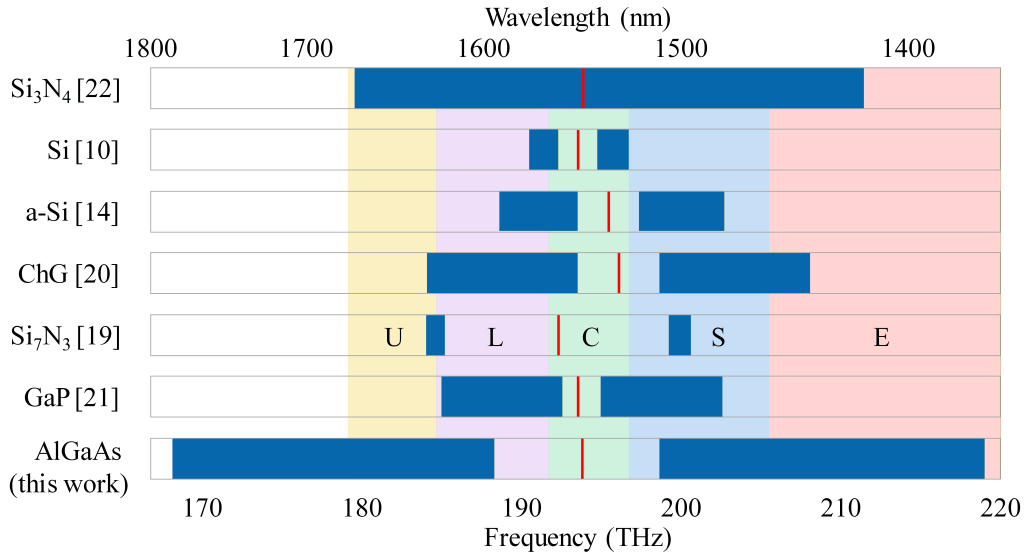}
\captionsetup{labelformat=default,labelsep=period} 
\caption{Net on-chip gain bandwidth of integrated optical parametric amplifiers pumped in the telecom band. Red line: pump wavelength; blue region: reported spectral range where net on-chip gain is positive. The E-, S-, C-, L-, and U-bands are all WDM channels.}
\label{figure5}
\end{figure}
For comparison, we list the parameters of representative integrated optical parametric amplifiers pumped in the telecom band in Table~\ref{Table1}. Notably, these optical parametric amplifiers are all based on nanoscale waveguides, and the nonlinear medium is only the waveguide core. In general, the net on-chip gain and bandwidth are key parameters for evaluating the performance of optical parametric amplifiers. Here, the net on-chip gain includes the influence of waveguide propagation loss, and the bandwidth represents the transparency window over which the net on-chip gain is positive.

While large on--off gain (70 dB \cite{torounidis2006fiber}) and bandwidth (26.19 THz \cite{ho2001200}) can be achieved in fibers, meter-long systems occupy substantial space. Benefiting from advances in nanoscale fabrication, integrated waveguides significantly reduce the footprint. Moreover, with a smaller mode area, the effective nonlinear coefficient is enhanced by orders of magnitude. Owing to the high nonlinearity of nanoscale waveguides, an on--off gain of up to 58.4 dB is obtained with millimeter-long AlGaAsOI waveguides, comparable with that of fiber systems. A 56.2-dB net on-chip gain is also demonstrated, enabled by the optimized propagation loss. Among these integrated platforms, $\rm {Si_7}N_3$ possesses a large nonlinear refractive index; however, the large fiber-to-chip coupling loss ($7\sim10.5$ dB) limits the on-chip pump power. Meanwhile, the severe propagation loss ($4.5\sim24$ dB/cm) reduces the net on-chip gain \cite{sahin2020optical,ooi2017pushing}. By optimizing the propagation loss to 1.4 dB/m and significantly increasing the waveguide length to 1.4 m on the $\rm {Si_3}N_4$ platform, the parametric gain is sufficiently enhanced to demonstrate net on-chip gain using a CW pump \cite{ye2021overcoming}. The bandwidth is further increased to 32.7 THz at the expense of only a 1-dB net on-chip gain by using a shorter waveguide \cite{zhao2025ultra}. Broader bandwidth is also achieved in the GaP platform, with higher nonlinearity, by reducing the waveguide length to 55.5 mm \cite{kuznetsov2025ultra}. Additionally, strong GVD limits the amplification bandwidth. The large index contrast of AlGaAsOI is conducive to dispersion engineering, enabling a small absolute GVD at the pump wavelength. With careful dispersion engineering, the GVD is reduced to $-27.2~\rm ps^2/km$ at 1550 nm. Figure~\ref{figure5} shows the net on-chip gain bandwidths of the parametric amplifiers in Table~\ref{Table1}, where the red line represents the pump wavelength and the blue region indicates the reported spectral range with positive net on-chip gain. Clearly, the AlGaAsOI platform achieves the largest net on-chip gain bandwidth, exceeding 415 nm. It spans 1370--1785 nm, covering the entire E-, S-, C-, L-, and U-bands used in wavelength-division multiplexing (WDM) systems. However, high-power operation in AlGaAsOI may be constrained by surface-state-induced nonlinear loss \cite{Kamel2023SurfaceWaveguides}, which limits efficient gain scaling and underscores the importance of surface passivation of such high-confinement waveguides \cite{guha2017surface}.

To further extend the bandwidth, a tapered waveguide can be employed to gradually vary the cross-section, thereby continuously modifying the local dispersion profile. This shifts the phase-matching condition and parametric gain peak along the propagation, enabling different spectral components to be amplified at different positions. As a result, the overall amplification bandwidth is broadened, albeit with reduced peak gain due to the absence of optimal phase matching at a single frequency across the entire length, as demonstrated in fiber-based systems \cite{inoue1994arrangement,su2000all,provino2002broadband}. In addition, higher-order dispersion can be implemented to increase and flatten the amplification bandwidth \cite{pu2018integrated,marhic1996broadband}, and can also induce an additional gain peak via discrete phase matching \cite{liu2012bridging,kuyken201150}. To further enhance the parametric gain, resonant designs such as microresonators \cite{kim2024parity} and photonic crystal waveguides \cite{sahin2020optical, ye2025boosting} can also be employed.

\section{Conclusion} 
Optical parametric amplification is investigated in AlGaAsOI waveguides with high nonlinearity and low loss. Using a picosecond pulsed source combined with a tunable CW signal, on--off signal parametric gain and idler translation gain of up to 56.9 dB and 58.4 dB are demonstrated. Moreover, benefiting from the low propagation loss of the device, the net on-chip gains for the signal and idler are as large as 54.7 dB and 56.2 dB, respectively. In addition, dispersion engineering increases the net on-chip gain bandwidth to exceed 415 nm. To the best of our knowledge, these results represent the largest gain and bandwidth in the telecom region to date, validating the outstanding performance of the AlGaAsOI platform for optical amplification. This work offers valuable guidance for the future design and tailoring of parametric amplifiers. Moreover, it provides a versatile platform for nonlinear optical processing applications within telecommunication systems, where large gain over a wide bandwidth is needed.



\medskip

\medskip
\textbf{Acknowledgements} \par 
This work is supported by European Research Council (REFOCUS 853522), Danish National Research Foundation, SPOC (ref. DNRF123), Innovationsfonden (INCOM 8057-00059B), and the Marie Sklodowska-Curie Grant (QUDOT-TECH 861097).

\medskip

%
\bibliographystyle{MSP}
\bibliography{mybib}

\end{document}